# PMU based Monitoring and Mitigation of Delayed Voltage Recovery using Admittances

Amarsagar Reddy Ramapuram Matavalam, *Student Member, IEEE*, and Venkataramana Ajjarapu, *Fellow, IEEE.*

*Abstract*—This paper analyzes the delayed voltage recovery phenomenon by simplifying the Western Electricity Coordinating Council (WECC) composite load model using the load admittance and the thermal relay dynamics. From this analysis, a closed form expression approximating the recovery time is derived and the key load parameters impacting the behavior of voltage recovery are identified. A monitoring scheme based on the measured load admittance is then proposed to detect the onset of stalling, even in the presence of voltage oscillations, and estimate the duration of the delayed voltage recovery. A mitigation scheme utilizing smart thermostats and offline learning is also derived to ensure that the voltage recovers to the pre-contingency voltage within a specified time. Both the monitoring and mitigation schemes only need local measurements at a substation making them promising for online applications. Results for the monitoring and mitigation schemes are described in detail for the IEEE 162 bus system validating the various assumptions used for the analysis and establishing the connection between the delayed voltage recovery phenomenon and load admittance.

## I. Nomenclature

Notation: Bold signifies a complex quantity. A subscript 'pre' ('post') for any of the following indicates the value just before (after) the contingency.

| | |
|---|---|
| $V_i$ | Internal voltage magnitude of composite load |
| $V_0$ | Substation voltage magnitude of composite load |
| $\theta_1$ | Thermal relay temperature to initiate tripping |
| $\theta_2$ | Thermal relay temperature to complete tripping |
| $T_{th}$ | Thermal relay time constant |
| $f_{th}$ | Motor-D fraction connected after thermal disconnection |
| $f_x$ | Fraction of x ∈ {Motor A, Motor B, Motor C, Motor D, Elec. Load, Static Load} in the composite load |
| $S_x$ | Apparent Power of x ∈ {Motor A, Motor B, Motor C, Motor D, Elec. Load, Static Load}. $S_x = P_x + j \cdot Q_x$ |
| $Y_x$ | Admittance of x ∈ {Motor A, Motor B, Motor C, Motor D, Elec. Load, Static Load}. $Y_x = G_x - j \cdot B_x$ |
| $Y_{ABCES}$ | Total admittance of A,B,C motors, Elec. & Static Load |
| $Y_{fd}$ | Admittance of Feeder in the composite load |
| $Y_{stall}$ | Stall admittance of Motor-D (1-$\phi$ motor). $Y_{stall} = G_{stall} - j \cdot B_{stall}$ |

## II. Introduction

IN recent years, deregulation of the markets giving priority to economic operations, there has been an increasing push to utilize the electric grid infrastructure to the best extent possible. This transformation in grid operations is causing the operators to operate the grid in more stressed conditions than usual, making it more likely for the problems to be manifested in the grid. One such problem is the phenomenon of short-term voltage instability which occurs mainly due to the stalling of induction motor loads, and can manifest in the form of fast voltage collapse or delayed voltage recovery.

During Fault Induced Delayed Voltage Recovery (FIDVR), the recovery of the voltage after a disturbance is delayed resulting in sustained low voltages for several seconds (>10 sec) [1]-[2]. FIDVR is mainly caused in systems with a moderate amount of single phase induction motor (IM) loads. After a large disturbance (fault, etc.), these motors, that are connected to mechanical loads with constant torque, stall and draw ~6 times their nominal current and leading to the depression of the system voltage for several seconds. There have been two kinds of solutions in literature to mitigate the FIDVR phenomenon – supply side methods (injection of dynamic VARs via SVC, etc.) and demand side methods (disconnection of loads using measurements, etc.).

Utilities usually employ the supply side solution by determining the amount and location of the SVCs and STATCOMs during the planning phase [3]-[6]. These methods use contingency sets and possible operating scenarios along with extensive time domain simulations to ensure that the installed devices can mitigate FIDVR over a wide range of contingencies. However, the planning phase cannot account for all rare & extreme events and so a measurement based method that can monitor FIDVR will aid the grid for rare/unexpected contingencies that were not planned.

The widespread adoption of Phasor Measurement Units (PMUs) by utilities has led to the development of methods to estimate the severity of FIDVR in an online manner to take appropriate control actions to prevent further stalling [7]-[11]. In [7] & [9], under-voltage (UV) load shedding scheme is proposed based on slope of voltage recovery and estimated recovery time. In [8], an MVA-Volt index is proposed to predict fault-induced low voltage problems on transmission systems. In [10], an online load shedding strategy by estimating the motor kinetic energy using PMUs is proposed to mitigate FIDVR. In [11], an index based on active and reactive power measurements is proposed to identify the effective load shedding locations.

A. R. Ramapuram Matavalam and V. Ajjarapu are with the Department of Electrical and Computer Engineering, Iowa State University, Ames, IA 50011 USA. (e-mail: amar@iastate.edu & vajjarap@iastate.edu).

All these above methods utilize the dynamic model of the 3-$\phi$ IM which is not the true cause of the FIDVR phenomenon as the stalling of 1-$\phi$ IM is the recognized cause of the FIDVR. In order to study the FIDVR phenomenon, the Composite Load Model (CMLD) has been developed by WECC [12] [13]. Analyzing the FIDVR using the CMLD model will enable better monitoring and control schemes to mitigate FIDVR. This model is complex and so a simplification of the dynamics is necessary for analysis. Recently [14], a new simplified dynamic model of motor stalling was proposed that incorporates switching using the gradient of an energy-like function. While this model exhibits the FIDVR phenomenon, the structure is completely different from the CMLD model and so it is hard to estimate these parameters in practice. In contrast, the present method retains the structure of the CMLD model, enabling us to leverage existing parameter estimation efforts [15] [16].

The structure of the paper is as follows: Section III presents a simplified analysis of the delayed voltage recovery phenomenon using the load admittance and discusses methods to estimate the time to recovery by using local PMU measurements at substations. Section IV validates the simplifying assumptions and discusses results for the IEEE 162 bus system. Section V presents the mitigation scheme utilizing smart thermostats along with the numerical results for the IEEE 162 bus system. Section VI concludes the paper along with possible research directions to be explored in the future.

### III. ANALYSIS OF DELAYED VOLTAGE RECOVERY USING LOAD ADMITTANCE

The composite load (CMLD) model essentially aggregates the various sub-transmission loads into three 3-$\phi$ induction motors (IM), (motors A, B & C), and an aggregate 1-$\phi$ IM (also referred as motor-D), representing the residential air conditioner (AC) loads. The overall structure of the composite load model is shown in Fig. 1 [12]-[13].

The 3-$\phi$ motors are represented using their equivalent circuit and the dynamics of these motors is usually in the order of 1-2 seconds. The 1-$\phi$ IM is the main reason why the FIDVR is observed. The 1-$\phi$ IM model represents the AC compressor motor, thermal relay, and contactors. Depending on the input voltage, the motor operates either in 'running' or 'stalled' state. The behavior of the motor as a function of the voltage can be understood from Fig. 2 which plots the active and reactive power demand as a function of the voltage for the normal operation (blue) and stalled operation (red) [12].

It can be seen that in the stalled state, the active power demand is ~3 times the nominal amount and the reactive demand is ~6 times the nominal amount compared to the normal 'running' state. The stalled 1-$\phi$ IM is represented as an admittance after stalling and as the terminal voltage increases, the active and reactive power drawn by the stalled IM increases in a quadratic manner. This large increase in the reactive power demand is the reason why the voltage at the load drops during stalling. The power demand is naturally reduced via thermal protection and takes around 5-15 seconds to operate which is the time the FIDVR phenomenon lasts. Despite the recovery, the concern is that the sustained low voltages can lead to events such as generator exciters reaching limits that initiate cascading, steering the system towards a blackout.

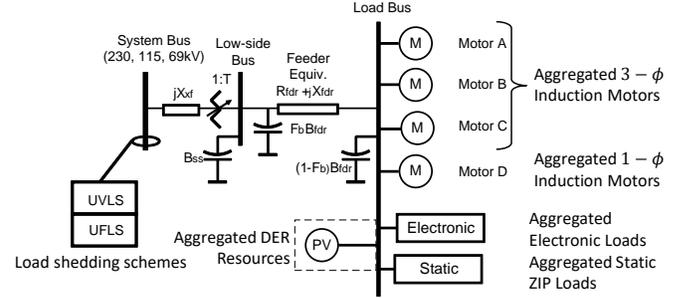

Fig. 1. Structure of the Composite Load Model [12]

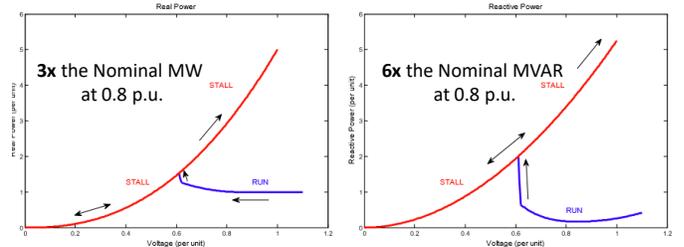

Fig. 2. Active (left) and reactive (right) power of 1-$\phi$ IM versus voltage [12]

As the thermal relay dynamics is much slower compared to the dynamics of the 3-$\phi$ IM, the fast dynamics of the 3-$\phi$ IM can be neglected and only the dynamics of the 1-$\phi$ IM thermal relay determines the overall behavior of the FIDVR phenomenon. Since the 1-$\phi$ IM are represented as an admittance after stalling, the 3-$\phi$ IM and the static loads can also be represented as a voltage dependent admittance. These observations and modelling assumptions lead to the admittance based representation of the composite load model.

#### A. Comparison of the behavior of load admittance and voltage during delayed voltage recovery

As a demonstration that the load admittance can indeed capture the load behavior during FIDVR, Fig. 3 plots the voltages and Fig. 4 plots the load conductance (real component of the admittance) for a normal, moderately severe (30% motor stalling) and very severe (60% motor stalling) delayed voltage recovery after a disturbance. The first observation is the voltage waveforms for both normal recovery and delayed recovery have oscillations due to the behavior of the other components in the system. In comparison, the conductance waveform is much better behaved for the normal recovery and delayed recovery. The oscillations in the voltage are due to the dynamic behavior of the external system (e.g. the generator exciter) and so the impact of these oscillations in the conductance are minimal as the oscillations do not impact the load behavior.

The next observation is that the voltage immediately after the fault is lower for higher amount of motor stalling. Similarly, the load conductance after the fault is cleared increases as the percent of motor stalling increases. However, it is not easy to quantify the severity of the FIDVR event from the voltages as the drop in voltage is not easily related to the severity and depends on the external network parameters. In contrast, the conductance makes it easy to quantify the severity of the event as the conductance increases in a nearly linear manner to the amount of motors stalled. Thus, it provides a quick way to characterize the severity of the FIDVR and enables monitoring and control schemes based on this quantification. The conductance during normal recovery quickly (< 1 sec) returns to the pre-contingency value. On the other hand, the conductance of the delayed voltage scenario has a sudden rise due to the stalling of the 1-ϕ IMs. The sudden rise can be used as a reliable indicator of the onset of the FIDVR phenomenon. The same cannot be said for the voltage as a severe FIDVR on a bus will depress voltages in neighboring buses even if there is no stalling in the neighboring buses.

Finally, the conductance for the delayed voltage scenario can be split into two parts – a flat region and a monotonically decreasing region. The flat region corresponds to the time to initiate the thermal tripping of 1-ϕ IM ($t_1$) and the region where the conductance reduces which corresponds to the time taken to complete the thermal tripping of 1-ϕ IM ($t_2$). It is much easier to distinguish between these phases of operation from the conductance plots compared to the voltage plots as the oscillations and other phenomenon can mask the exact time of transition.

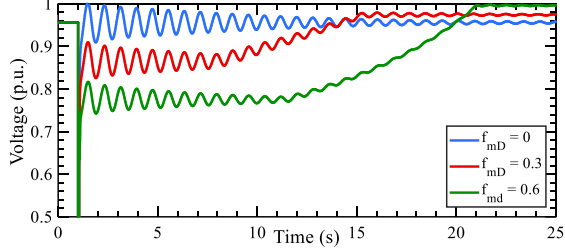

Fig. 3. Voltage response with various motor stalling proportion

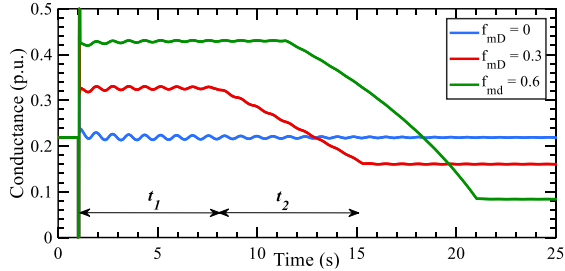

Fig. 4. Load conductance with various motor stalling proportion. $t_1$ & $t_2$ are indicated for $f_{mD} = 0.3$.

The load susceptance has a similar behavior as the load conductance for the FIDVR scenario and in the rest of the paper the susceptance is not plotted in the interest of space. By observing various conductance (susceptance) plots for various proportions of stalled motor, two observations can be made: (1) the load conductance (susceptance) is nearly constant till the motor thermal protection triggers (2) the slope of conductance (susceptance) due to the thermal disconnection is almost constant. Based on these observations, the following hypothesis is suggested - by measuring the admittance just after the FIDVR begins, the values of $t_1$ and $t_2$ can be estimated from the load parameters using the admittance based model.

Fig. 5 shows the structure of the admittance based composite load model with the load connected to a generator with voltage E and a transmission line with admittance $\mathbf{Y}_{trans}$. The equivalent feeder admittance is denoted by $\mathbf{Y}_{fd}$, and includes the substation tap transformer and the shunt compensation to compensate for the voltage drop in the feeder. The PMU is present at the substation before $\mathbf{Y}_{fd}$ and measures $V_0$ & load current which can be used to calculate the internal voltage from $\mathbf{Y}_{fd}$. Thus, in the derivations and results, we assume that we know the value of $V_i$. The A, B & C motors, electronic loads and static loads are represented by $\mathbf{Y}_{ABCES(V_i)}$ and the 1-ϕ motor is represented by $\mathbf{Y}_{mD}$. The admittance $\mathbf{Y}_{ABCES(V_i)}$ is a function of the voltage in order to account for the dynamics of the A, B & C motors and is not constant with time. After a severe fault, the stalled 1-ϕ IM admittance is given by $\mathbf{Y}_{stall}$. The fraction of 1-ϕ IM connected after stalling is determined by the fraction $f_{th}$ which is the output of the thermal relay. The thermal relay block diagram is shown in Fig. 6, where the thermal power dissipated in the motor given by $P_{th}$ (equal to $V_i^2 \cdot G_{stall}$), $T_{th}$ is the thermal relay time constant and $\theta$ is the motor temperature estimated by the relay [12]. Analysis of this simplified model along with the thermal relay dynamics is discussed next to estimate times $t_1$ & $t_2$ from measurements.

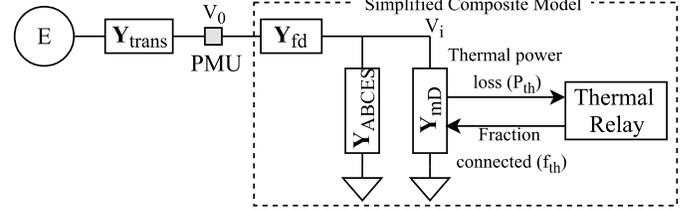

Fig. 5. Two bus system with the simplified composite load model

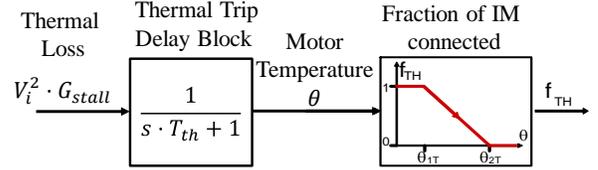

Fig. 6. The thermal relay dynamics of the 1-ϕ IM [12]

### B. Time taken to initiate motor disconnection ($t_1$)

The general expression for the voltage $V_i$ is given by (1) with $\mathbf{Y}_{eff}$ denoting the effective admittance seen by the motors. After the motor stalling of 1-ϕ IM, the post-contingency voltage is given by (2) with $\mathbf{Y}_{mD}$ replaced by $\mathbf{Y}_{stall}$. Initially the internal temperature is zero and the thermal loss is zero. As the stalling condition occurs suddenly, the input to the thermal delay block can be approximated by a step function with height given by $P_{th_{post}}$. This step input passes through a 1st order time delay block with a delay $T_{TH}$ and the internal temperature increases exponentially as shown in (4). From the thermal relay dynamics, the $f_{th}$ fraction output remains equal to 1 till the temperature reaches $\theta_1$. Thus $G_{stall}$ remains same and the time taken for the temperature to reach $\theta_1$ can be calculated by substituting (3) in (4) to get (5).

$$V_i^2 = \frac{E^2 \cdot |\mathbf{Y}_{eff}|^2}{|\mathbf{Y}_{eff} + \mathbf{Y}_{ABCES(V_i)} + \mathbf{Y}_{mD}|^2}; \mathbf{Y}_{eff} = \frac{\mathbf{Y}_{fd} \cdot \mathbf{Y}_{trans}}{\mathbf{Y}_{fd} + \mathbf{Y}_{trans}} \quad (1)$$

$$V_{i_{post}}^2 = \frac{E^2 \cdot |\mathbf{Y}_{eff}|^2}{|\mathbf{Y}_{eff} + \mathbf{Y}_{ABCES(V_i)} + \mathbf{Y}_{stall}|^2} \quad (2)$$

$$P_{th_{post}} = V_{i_{post}}^2 \cdot G_{stall} \quad (3)$$

$$\theta_1 = P_{th_{post}} \left(1 - e^{(-t_1/T_{th})}\right) \quad (4)$$

$$t_1 = -T_{th} \cdot \ln\left(1 - \theta_1 / \left(V_{i_{post}}^2 \cdot G_{stall}\right)\right) \quad (5)$$

Thus, the various load parameters can be used to estimate the time taken for the 1-ϕ IM to start disconnecting. Next, we can determine the time taken for the motor temperature to rise from $\theta_1$ to $\theta_2$ by understanding how the thermal trip fraction $f_{TH}$ varies with time.

### C. Time taken to complete motor disconnection ($t_2$)

After the thermal disconnection begins, the voltage ($V_{i(f_{th})}$) and thermal power dissipated ($P_{th(f_{th})}$) depend on the fraction of the motors still connected and are given by (6)-(7). The fraction $f_{th}$ multiplies the stall admittance in (6), reducing the denominator magnitude and increasing the voltage. (8) & (9) follow from the relation between $P_{th}$, $\theta$ and fraction $f_{th}$ from the block diagram in Fig. 6. Differentiating (9) on both sides and using (7) & (8) leads to (10).

$$V_{i\,(f_{th})}^2 = \frac{E^2 \cdot |Y_{eff}|^2}{|Y_{eff} + Y_{ABCES(V_i)} + f_{th} \cdot Y_{stall}|^2} \quad (6)$$

$$P_{th(f_{th})} = V_{i\,(f_{th})}^2 \cdot G_{stall} \quad (7)$$

$$\frac{d\theta}{dt} = \frac{1}{T_{Th}}(P_{th} - \theta) \quad (8)$$

$$\theta = (\theta_2 - \theta_1) \cdot (1 - f_{TH}) + \theta_1 \quad (9)$$

$$\frac{df_{th}}{dt} = \frac{-d\theta/dt}{(\theta_2 - \theta_1)} = \frac{\left(\theta_2 - (\theta_2 - \theta_1) \cdot f_{th} - V_{i\,(f_{th})}^2 \cdot G_{stall}\right)}{T_{th}(\theta_2 - \theta_1)} \quad (10)$$

Equation (10) will enable us to understand the behavior of the thermal relay and estimate the time taken to disconnect the 1-$\phi$ IM. Initially, the value of the voltage is post contingency voltage ($V_{i_{post}}$), the value of $f_{th}$ is 1 and the value of the slope $df_{th}/dt$ is negative implying that the $f_{th}$ will reduce from 1 and the voltage will increase in magnitude. As the voltage increases and the $f_{th}$ decreases, the slope becomes further negative and increases the rate of rise of voltage. Finally, as the value of $f_{th}$ reaches 0, the voltage reaches close to the pre-contingency voltage ($V_{i_{pre}}$) at which time all the 1-$\phi$ IM are disconnected and the thermal trip relay stops operating. In reality the voltage after the system recovers from FIDVR is slightly higher than the initial pre-contingency voltage but this difference is small.

The differential equation (10) is non-linear which can be solved numerically but has no analytical solution, making it hard to analyze for various scenarios. The observation that the slope of the load admittance is nearly constant during the disconnection phase suggests that we can use the mean of the slope at the two extreme values of $f_{th}$ to approximate $df_{th}/dt$. This approximation is used to derive $t_{2-approx}$ in the equations below. The initial slope $df_{th}/dt$ uses $f_{th} = 1$ while the final slope $df_{th}/dt$ uses $f_{th} = 0$. The voltages $V_{i_{post}}$ and $V_{i_{pre}}$ are used in the expressions for the initial and finals slopes in (11) & (12). The change in fraction ($\Delta f_{th}$) is -1 as the fraction goes from 1 to 0 and dividing $\Delta f_{th}$ by the slope gives the approximate time to complete disconnection given in (15).

$$\left.\frac{df_{th}}{dt}\right|_{initial\,(f_{th}=1)} = \frac{1}{T_{th}(\theta_2 - \theta_1)}\left(\theta_1 - V_{i_{post}}^2 G_{stall}\right) \quad (11)$$

$$\left.\frac{df_{th}}{dt}\right|_{final\,(f_{th}=0)} = \frac{1}{T_{th}(\theta_2 - \theta_1)}\left(\theta_2 - V_{i_{pre}}^2 G_{stall}\right) \quad (12)$$

$$\left.\frac{df_{th}}{dt}\right|_{mean} = \frac{1}{2}\left(\left.\frac{df_{th}}{dt}\right|_{initial} + \left.\frac{df_{th}}{dt}\right|_{final}\right) \quad (13)$$

$$\left.\frac{df_{th}}{dt}\right|_{mean} = \frac{\left(\theta_1 + \theta_2 - \left(V_{i_{pre}}^2 + V_{i_{post}}^2\right)G_{stall}\right)}{2T_{th}(\theta_2 - \theta_1)} \quad (14)$$

$$t_{2-approx} = \frac{\Delta f_{th}}{df_{th}/dt} = \frac{2T_{th}(\theta_2 - \theta_1)}{\left(\left(V_{i_{pre}}^2 + V_{i_{post}}^2\right)G_{stall} - \theta_1 - \theta_2\right)} \quad (15)$$

Thus, utilizing the voltage measurements $V_{i_{pre}}^2$ & $V_{i_{post}}^2$ from the PMU along with the load parameters, the time to recover from FIDVR can be determined, which is a quantification of the FIDVR severity. The stall conductance depends on the proportion of stalled load and needs to be estimated from the measurements given some basic information about the load composition from the utility.

### D. Estimating stall conductance from measurements

To determine the stall conductance, we exploit the fact that the behavior of the A,B,C motors, electronic load and static load do not change too much from their pre-contingency values. There have been studies that provide the composition of the load at a particular region in terms of the A,B,C motors, electronic loads and the static ZIP parameters [15]. These percentages are used to separate the total conductance into individual conductances at normal operation as follows.

$$P_{load} = P_{ABCE} + P_{st} \cdot (V^2 \cdot f_{st_Z} + V \cdot f_{st_I} + f_{st_P}) + P_{mD} \quad (16)$$

$$P_{ABCE} = (f_{mA} + f_{mB} + f_{mC} + f_{Elec}) \cdot P_{load} \;;\; P_{st} = f_{st} \cdot P_{load} \quad (17)$$

$$G_{load} = \frac{P_{ABCE}}{V^2} + P_{st} \cdot \left(f_{st_Z} + \frac{f_{st_I}}{V} + \frac{f_{st_P}}{V^2}\right) + G_{mD} \quad (18)$$

Assuming that the active power of A, B & C motors and electronic loads do not change significantly after the fault, the following equation for the post-contingency load conductance can be written in terms of pre-contingency powers.

$$G_{load_{post}} = \frac{P_{ABCE_{pre}}}{V_{post}^2} + P_{st_{pre}} \cdot \left(f_{st_Z} + \frac{f_{st_I}}{V_{post}} + \frac{f_{st_P}}{V_{post}^2}\right) + G_{stall} \quad (19)$$

Using (17) to express $P_{ABCE_{pre}}$ & $P_{st_{pre}}$ in (19), the stall conductance of the composite load can be expressed as (20).

$$G_{stall} = G_{load_{post}} - \frac{P_{load_{pre}}}{V_{post}^2}\left(f_A + f_B + f_C + f_E + f_{st} \cdot f_{st_P} \right.$$
$$\left. + f_{st} \cdot f_{st_I} \cdot V_{post} + f_{st} \cdot f_{st_Z} \cdot V_{post}^2\right) \quad (20)$$

Thus, using the fraction of load split between the A,B,C motors, electronic loads and the static ZIP parameters, the stall conductance can be calculated using load voltage, load conductance and load power which can be measured at a substation. Similar equations can be written for estimating the stall susceptance using load susceptance and reactive power. In case the A,B,C motors were installed with UV relays, their power would be altered after the FIDVR event and this can be included in the expression above.

It can be observed that the key parameters that impact the recovery time from (5) and (15) are $G_{stall}, B_{stall}, \theta_1, \theta_2, \& T_{th}$. Also, as $V_i$ is estimated from $V_0$, $Y_{fd}$ is important. These parameters have been shown to be the most crucial parameters that impact the behavior of FIDVR using simulations on actual utility data [17]. The derivation (1)-(15) is from physical principles and provides a theoretical basis for these parameters to be the most critical parameters determining the behavior of the FIDVR response. An interesting observation is that even though a simplified equivalent of the system is used to derive the equations, the terms $E$ & $Y_{trans}$ do not appear in the final expressions as their impact is indirectly present in the voltage measurements. Thus, there is no need to estimate $E$ & $Y_{trans}$ for predicting the recovery time. Fig. 7 summarizes the proposed methodology for detection and monitoring of FIDVR using admittances as a flow chart. Next results validating the various assumptions and the demonstrating the accuracy of the proposed methodology are discussed using the IEEE 162 bus system.

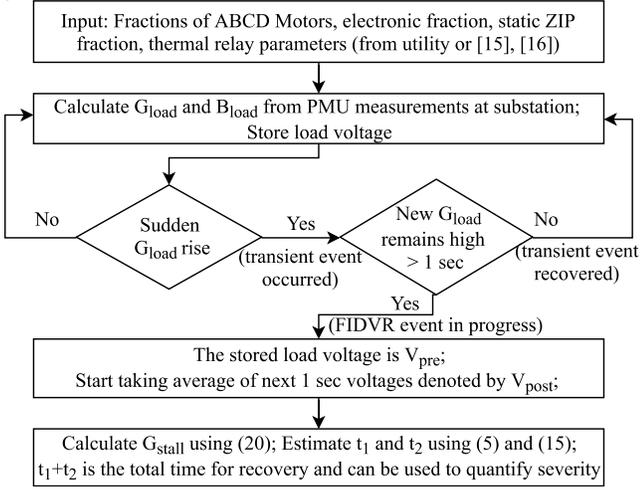

Fig. 7. Flowchart for detecting and monitoring FIDVR using measurements

## IV. NUMERICAL RESULTS

Numerical examples demonstrating the accuracy of the prediction of recovery time are presented in this section. Simulations are done in PSSE on the IEEE 162 bus system with loads that have been converted to composite load models. More information about the system is described in [6] [18]. The parameters of the composite load model are $f_{mA} = 0.15, f_{mB} = 0.05, f_{mC} = 0, f_{Elec} = 0.15, T_{th} = 15s, \theta_1 = 0.9 \& \theta_2 = 1.5$. The $f_{mD}$ is increased from 10% to 45% and $f_{stat}$ is reduced from 55% to 20% in steps of 5% giving 8 cases. At 1 second, a 3-phase to ground fault is applied at bus 120 for 3 cycles (50 ms) and this leads to FIDVR being observed at load buses with 1-$\phi$ IM. In the interest of space the behavior of the load at bus 135 is analyzed as similar results were observed at other load buses with 1-$\phi$ IM. The behavior of composite loads at other buses is similar and is omitted in the interest of space. The voltage and the load conductance for the various scenarios is plotted in Fig. 8 and Fig. 9.

For all the scenarios, the load voltage oscillates making it is hard to detect motor stalling just from the voltages, especially for the cases with small percentage of 1-$\phi$ IM. Also, the oscillations make it hard to quantify the severity of the FIDVR event as the distance between the voltages get closer to each other with the same proportion of the increment in the proportion of the 1-$\phi$ IM. Furthermore, it is hard to detect the instant when the thermal relay begins disconnection just using the voltage information. In contrast, the stalling can be clearly detected by the sudden rise of load conductance after the fault and the less amount of oscillations in the conductances make it easy to quantify the severity of an event. The severity of the FIDVR is calculated based on the amount of conductance rise from the base case as the distance between the conductances is nearly constant with the same increment in the proportion of the 1-$\phi$ IM. Additionally, it is easy to identify the instant when the disconnection of the 1-$\phi$ IM begins as it is the instant when the conductance plot starts to decrease. Thus, the behavior of the conductance is easier to analyze in order to understand the FIDVR phenomenon for larger systems as well.

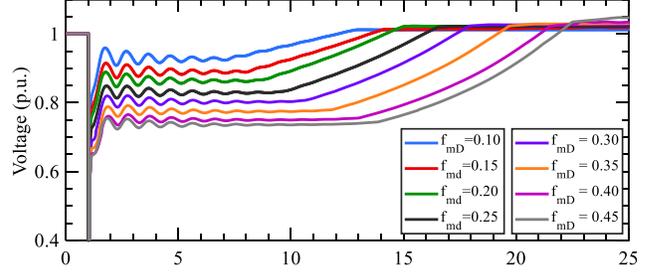

Fig. 8. Voltage at Bus 135 for $f_{mD} = 0.10$ to $f_{mD} = 0.45$ with fault at 1s

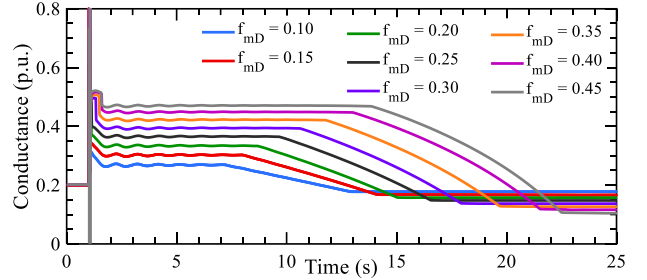

Fig. 9. Load conductance for $f_{mD} = 0.10$ to $f_{mD} = 0.45$ with fault at 1s

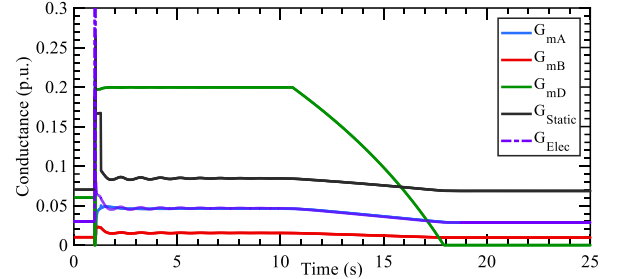

Fig. 10. Conductance of the components of the CMLD model for $f_{mD} = 0.3$

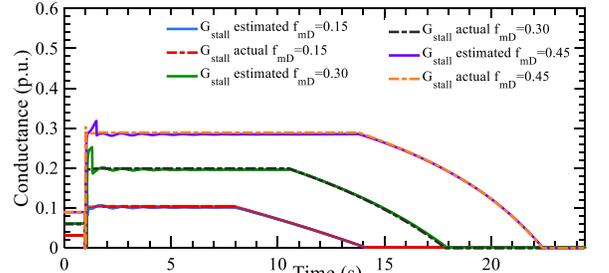

Fig. 11. Estimated and actual stall conductance for $f_{mD} = 0.15, 0.3 \& 0.45$

In order to validate the assumption that the behavior of the 1-$\phi$ IM dominate the overall load behavior, the conductances of the individual A,B,C,D motors, electronic and static load is plotted versus time in Fig. 10 for the case with $f_{mD} = 30\%$. The conductances before the fault are approximately in the same proportion to their powers as the voltages are close to 1 p.u. After the FIDVR event, the conductance of the 1-$\phi$ IM rises to several times its nominal value while the conductance of the other components changes by a much smaller amount. Thus, the increment of the load conductance is mainly due to the change of the conductance of the 1-$\phi$ IM due to the stalling. This validates the key proposition that the time to recover from FIDVR is mainly dependent on the behavior of the 1-$\phi$ IM. Next, to verify that the expression in (20) can indeed estimate the stalled motor conductance, the estimated 1-ϕ IM conductance along with the actual 1-ϕ IM conductance is plotted versus time in Fig. 11 for three different proportions of 1-ϕ IM. It can be observed that the estimated motor conductance has some switching transients and oscillations due to the behavior of the other components of the composite load. But after a short time (~1 sec), the estimated conductance matches with the actual 1-ϕ IM conductance. Even though we are only interested in the 1-ϕ IM conductance just after the FIDVR event begins, the expression in (20) is able to estimate the conductance over the entire FIDVR event, validating (20).

Now that the simplifying assumptions of the model have been validated, the times $t_1$ and $t_2$ that characterize the FIDVR event can be estimated from the equations (5) and (15) using the estimated stall conductance and the measured pre-contingency and post-contingency voltage. As there are oscillations present in the voltages, the average voltage between 1 sec & 2 sec after the fault is cleared is used for the post-contingency voltage. The results of the actual times along with the estimated times for the FIDVR with varying amounts of $f_{mD}$ is presented in Table I.

Table I. Actual and estimated $G_{stall}$, $t_1$ & $t_2$ with varying 1-ϕ IM proportion

| $f_{md}$ | $G_{stall}$ actual | $t_1$ actual | $t_2$ actual | $G_{stall}$ estim. | $t_1$ estim. | $t_2$ estim. |
| --- | --- | --- | --- | --- | --- | --- |
| 10% | 0.07 | 6.3 s | 5.6 s | 0.068 | 6.1 s | 5.3 s |
| 15% | 0.10 | 7.0 s | 6.1 s | 0.097 | 6.85 s | 5.9 s |
| 20% | 0.13 | 7.7 s | 6.3 s | 0.124 | 7.6 s | 6.3 s |
| 25% | 0.17 | 8.7 s | 6.8 s | 0.166 | 8.6 s | 6.9 s |
| 30% | 0.20 | 9.7 s | 7.2 s | 0.195 | 9.55 s | 7.3 s |
| 35% | 0.23 | 10.9 s | 7.7 s | 0.225 | 10.7 s | 7.9 s |
| 40% | 0.26 | 12.1 s | 8.4 s | 0.254 | 12.0 s | 8.5 s |
| 45% | 0.29 | 12.9 s | 8.7 s | 0.285 | 13.0 s | 8.9 s |

It can be seen that the estimated stall conductance and the estimated times $t_1$ and $t_2$ have an error less than 5% of the true measured values. The maximum error occurs when the $f_{mD}$ percent is less due to the fact that the impacts of the dynamics of A,B,C motors are comparable in these scenarios. As the $f_{mD}$ percent increases, the estimated times get closer to the actual values and start becoming more than the actual values, validating the various approximations made in the derivations of the expressions for recovery time. As the measurements are all at a single substation, the proposed method is a local method to detect and quantify FIDVR recovery time in an online manner.

The accuracy of the method is tested on several different proportions of A,B,C Motors and the maximum error was ~10% over all scenarios. We also tested the method in the case of a badly tuned generator exciter that caused sustained oscillations in the voltage oscillations. The proposed admittance based method accurately identified the FIDVR in presence of oscillations and was able to estimate the times with similar accuracy. These results are omitted in the interest of space. It can be seen from the conductance plot in Fig. 9 that the times $t_1$ and $t_2$ vary almost linearly with the load conductance rise. This relation was observed for all the scenarios with varying A,B,C motor proportions and this fact is utilized in the next section for the mitigation of FIDVR.

## V. Mitigation Schemes to Reduce Recovery Time

Now that we have a local method to estimate FIDVR recovery time, a control scheme based on the same framework is the next step. If the estimated recovery time is more than a specified time (determined by the utility), then a control scheme needs to be triggered that ensures that the voltage recovers faster. The main reason for mitigating FIDVR is that depressed voltages over several seconds can lead to generators reaching their field current limits as they try to supply the excessive reactive power required by the load [1]. At present, utilities mitigate FIDVR by installing SVC's using extensive simulations [5]. However, by using the PMU's to detect FIDVR in real-time, we can utilize the flexibility of the loads to mitigate FIDVR.

As the majority of the stalled motors are the residential air conditioners that are not equipped with the UV relays, disconnecting these motors by smart thermostats is used to recover from FIDVR. The increasing use of smart thermostats in modern residences enables the utility to utilize the thermal capacity of the residences to improve the system behavior. The smart thermostats can turn the AC's off quickly when they receive a signal. A stalled AC is not actually performing any useful work as it is not operating as usual and so disconnecting this motor will not impact the customer. To determine how much proportion of the AC load should be tripped, we utilize the approximate linear relation between the recovery time and the load conductance rise as it encodes the information of $E$ and $Y_{trans}$ and a derivation is shown next.

### A. Relation between recovery time and load conductance rise

The expression for $t_1$ in (5) can be simplified using the Taylor expansion $\ln(1-x) \approx -x$ and assuming $Y_{eff} \approx Y_{trans}$, leading to the following expression for $t_1$.

$$t_1 \approx \frac{(T_{th} \cdot \theta_1)}{V_{i_{post}}^2 \cdot G_{stall}} \; ; \; V_{i_{post}}^2 \approx \frac{E^2 \cdot |Y_{trans}|^2}{|Y_{trans} + Y_{load}|^2} \qquad (21)$$

$$t_1 \approx \frac{(T_{th} \cdot \theta_1)}{E^2 \cdot G_{stall}} \cdot \left|1 + \frac{Y_{load}}{Y_{trans}}\right|^2 \qquad (22)$$

$$t_1 \approx \frac{(T_{th} \cdot \theta_1)}{E^2 \cdot G_{stall}} \cdot \left|1 + \frac{Y_{load_{pre}}}{Y_{trans}} + \frac{\Delta Y_{load}}{Y_{trans}}\right|^2$$

Utilizing the fact that the load increment of conductance ($\Delta G_{load}$) and susceptance ($\Delta B_{load}$) at FIDVR are nearly the same, and using the Taylor expansion $|1 + x|^2 \approx (1 + 2 \cdot \mathcal{Real}(x))$, we get the following approximation for $t_1$

$$t_1 \approx \frac{(T_{th} \cdot \theta_1)}{E^2 \cdot G_{stall}} \cdot \left|1 + \frac{Y_{load_{pre}}}{Y_{trans}} + \frac{(1-i) \cdot \Delta G_{load}}{Y_{trans}}\right|^2 \qquad (23)$$

$$t_1 \approx \frac{(T_{th} \cdot \theta_1)}{E^2 \cdot G_{stall}} \cdot \left(1 + 2\mathcal{Real}\left(\frac{Y_{load_{pre}}}{Y_{trans}} + \frac{(1-i) \cdot \Delta G_{load}}{Y_{trans}}\right)\right) \qquad (24)$$

Since the values of $\theta_1, \theta_2$ and $G_{stall}$ are specified in a per unit basis, their ratio will not change based on the $f_{mD}$ proportion and so the only variable in the expression for $t_1$ that changes with the $f_{mD}$ is $\Delta G_{load}$. A similar derivation can be done for $t_2$ and we get the following expressions for $t_1$ and $t_2$.

$$t_1 \approx \alpha_0 \cdot \Delta G_{load} + \alpha_1; \; t_2 \approx \beta_0 \cdot \Delta G_{load} + \beta_1 \qquad (25)$$

In the above expression, $\alpha_0, \alpha_1, \beta_1$ & $\beta_2$ are functions of $E$ and $Y_{trans}$. A similar linear approximation using $\Delta B_{load}$ can also be derived which can be used when UV relays disconnect A, B & C motors. These linear coefficients can be learned from offline simulations and utilized in an online manner to estimate recovery time, without needing to estimate the equivalent $E$ and $Y_{trans}$. Initially, it might seem that different contingencies can lead to different values of $\alpha_0, \alpha_1, \beta_1$ & $\beta_2$ and so the amount of offline studies might be large. However, [6] demonstrates that most of the severe contingencies can be clustered together and their impact on the system can be captured by a few contingencies in each cluster. Thus, we need to perform offline studies only for these few contingencies and store the corresponding coefficients. This linear approximation can be used to estimate the impact of the AC disconnection on the recovery time and is discussed in the next sub-section.

### B. Impact of AC disconnection on recovery time

Let the initial rise in conductance at the beginning of FIDVR be $G_0$. At a time $\tau_0$ sec, $(1 - \gamma)$ fraction of the AC load is disconnected by the smart thermostats and the conductance immediately after the disconnection is $G_1 = \gamma G_0$. The conductance then remains constant for a further $\tau_1$ sec before the thermal disconnection begins. The total time for disconnection to begin is $t_1 = (\tau_1 + \tau_2)$.

From the previous sub-section, $t_1$ & $t_2$ can be approximated as linear functions of the conductance rise. As the time $t_1$ is determined by the heating of the motor coil, the average conductance of $G_0$ and $G_1$ weighted by $\tau_0$ and $\tau_1$ determines $t_1$. Similarly, as the time $t_2$ is determined by the amount of remaining conductance to disconnect, it depends only on $G_1$. The total recovery time for the FIDVR is given by $t_1 + t_2$ and this should be equal to $t_{sp}$, the time specified by the utility. The following equations follow from their definitions, using (25) along with (26) leads to (28)-(29).

$$G_{avg} = \frac{\tau_0 G_0 + \tau_1 G_1}{\tau_0 + \tau_1} = \frac{\tau_0 + \gamma \tau_1}{\tau_0 + \tau_1} G_0 \quad (26)$$

$$t_1 + t_2 = t_{sp} \; ; \; \tau_1 = t_1 - \tau_0 \quad (27)$$

$$t_1 = \alpha_0 \cdot G_{avg} + \alpha_1 = \alpha_0 \cdot \frac{\tau_0 + \gamma \tau_1}{\tau_0 + \tau_1} G_0 + \alpha_1 \quad (28)$$

$$t_2 = \beta_0 \cdot G_1 + \beta_1 = \beta_0 \cdot \gamma \cdot G_0 + \beta_1 \quad (29)$$

The utility would like to estimate the amount of AC's to trip at a particular time $\tau_0$ so that the total recovery time is equal to a specified time, $t_{sp}$. Thus, combining (27)-(29) and eliminating $\tau_1$, the quadratic equation (30) can be obtained. This can be solved analytically to estimate the value of $\gamma$ and only $0 < \gamma < 1$ is physically realizable.

$$\begin{aligned}(t_{sp} - \beta_1 - \beta_0 G_0 \gamma) \cdot (t_{sp} - \beta_1 - \alpha_1 - (\beta_0 + \alpha_0) G_0 \gamma) \\ = \alpha_0 \tau_0 G_0 (1 - \gamma)\end{aligned} \quad (30)$$

After solving (30) for $\gamma$, $(1 - \gamma)$ fraction of the AC load has to be tripped at $\tau_0$ time instant to ensure that the FIDVR event recovers within the specified time. To test this method the IEEE 162 bus system is used and multiple offline studies for a few representative contingencies identified in [18] (Appendix A) are performed with the varying proportion of motor A, B, C & D. These offline simulations are used to learn the $\alpha_0, \alpha_1, \beta_1$ & $\beta_2$ coefficients for the few representative contingencies and the results using these to determine the load disconnection is described next.

### C. Numerical Results on IEEE 162-Bus example

The linear approximation of the times $t_1$ and $t_2$ are determined from offline studies in PSSE on the IEEE 162 for the scenario of fault on Bus 120 for 3 cycles. The expressions for the time $t_1$ and $t_2$ in terms of the load conductance at Bus 135 is estimated to be (31).

$$t_1 \approx 39.5 \cdot \Delta G_{load} + 2.4; \; t_2 \approx 17.5 \cdot \Delta G_{load} + 4 \quad (31)$$

While estimating the linear relation from offline studies, the scenarios with larger $t_1$ and $t_2$ are given more weightage as we need more accuracy when the amount of FIDVR is high. To test the generalizability of (31), a fault at bus 75 is used, which is in the same contingency cluster (Appendix A - [18]) but is different from the trained scenario. The FIDVR recovery time is measured from the simulations and is compared with the estimated time from (31). The results are presented in Table II.

Table II. Actual and estimated $t_1$ and $t_2$ using (31)

| $f_{md}$ | $\Delta G_{load}$ | $t_1$ actual | $t_2$ actual | $t_1$ estim. | $t_2$ estim. |
|---|---|---|---|---|---|
| 10% | 0.07 | 6.3 s | 5.6 s | 5.2 s | 5.2 s |
| 15% | 0.1 | 7.0 s | 6.1 s | 6.4 s | 5.8 s |
| 20% | 0.13 | 7.7 s | 6.3 s | 7.5 s | 6.3 s |
| 25% | 0.16 | 8.7 s | 6.8 s | 8.7 s | 6.8 s |
| 30% | 0.19 | 9.7 s | 7.2 s | 9.9 s | 7.3 s |
| 35% | 0.22 | 10.9 s | 7.7 s | 11.1 s | 7.9 s |
| 40% | 0.245 | 12.1 s | 8.4 s | 12.1 s | 8.3 s |
| 45% | 0.27 | 12.9 s | 8.7 s | 13.1 s | 8.7 s |

It can be seen that the times estimated from the linear expressions for the cases with $f_{mD} > 20\%$ are within 0.25 sec of the true recovery time, demonstrating that the linear expressions for the bus are accurate for the severe FIDVR events. Next, we can determine the amount of smart thermostats to disconnect to ensure voltage recovery within a pre-specified time for severe FIDVR events. To demonstrate this, the scenario with $f_{md} = 30\%$ with the fault at bus 75 is chosen in which voltage recovers in 16.9 sec. The amount of ACs to disconnect is estimated using (30) for recovery time of 14 sec and 13 sec with a $\tau_0$ of 2 sec and 3 sec, leading to 4 test cases.

The results of the total time to recover for these cases are summarized in Table III and it can be seen that the error percent in the total recovery time for these cases is within 5%, validating the estimate of the percent of ACs to disconnect. It can also be observed that the percent of the ACs to disconnect increases as the specified time reduces and the $\tau_0$ increases, which is intuitive. The corresponding voltages plotted in Fig. 12 demonstrate that the voltages recover to the pre-contingency voltage in a controlled manner after the AC disconnection.

Table III. Specified and actual recovery time and corresponding disconnection

| $t_{sp}$ | $\tau_0$ | AC disconnect | $(t_1 + t_2)$ | Error (%) |
|---|---|---|---|---|
| 14 sec | 2 sec | 37 % | 13.45 s | -3.9 % |
| 14 sec | 3 sec | 40 % | 13.35 s | -4.6 % |
| 13 sec | 2 sec | 49 % | 12.70 s | -2.3 % |
| 13 sec | 3 sec | 54 % | 12.75 s | -1.7 % |

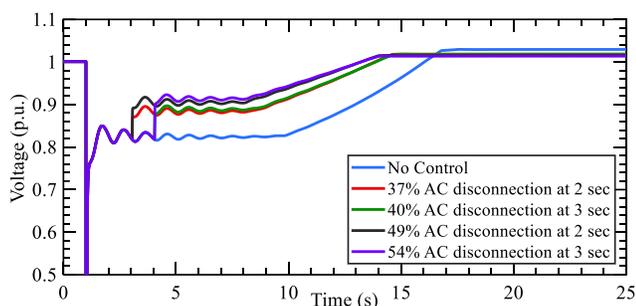

Fig. 12. Voltage at Bus 135 with and without load control

Thus, using the offline linear coefficients and the measured load conductance rise, the percent of ACs to disconnect for improving the load voltage can be estimated in an online manner at the substation and will enable the utilities to adhere to their specific transient voltage criteria [3].

## VI. CONCLUSION AND FUTURE STUDIES

In this paper, the FIDVR phenomenon is analyzed based on the load admittance by simplifying the composite load model using practical assumptions about the load behavior. The analysis simplified the load dynamics into a 1-D non-linear dynamical system by exploiting the thermal relay dynamics and facilitated the derivation of closed form expressions approximating the FIDVR recovery time. These expressions only use local PMU measurements and can be used to quantify the severity of the FIDVR based on a few load parameters. The assumptions and the methodology are verified on the IEEE 162 bus system with various contingencies, load parameters, and generator parameters. It was observed that the admittance based method accurately identified the onset of FIDVR, even in the presence of voltage oscillations; and the recovery time was within 10% of the true value for a variety of scenarios and load composition. A control scheme based on offline learning and utilizing smart thermostats to recover the voltage within a specified time is also proposed based on the admittance and is tested on the IEEE 162 bus system under various contingencies and the error is observed to be ~0.25 s for severe contingencies.

A natural extension of the proposed method is to use wide-area measurements along with the topology information and this will enable more targeted control to mitigate FIDVR. Data from distribution feeders [19] suggests that FIDVR occurs more frequently in distribution systems than the transmission systems. Thus, at present we are investigating how measurements from $\mu$PMUs can be used to localize the stalling and how the reactive support from the Distributed Energy Devices (solar, etc.) can be used to mitigate FIDVR [20]. One of the concerns of using PV inverters to correct FIDVR is the over voltage due to over compensation after the voltage recovery. The recovery time estimation can be used to ensure that control is only applied for the estimated recovery time and limits the over voltage duration after FIDVR. Traditionally, planning reactive support to mitigate FIDVR used the voltages to quantify the FIDVR phenomenon. The analysis presented in this paper suggests that load admittance is a more natural variable to analyze the FIDVR phenomenon and we will study how reactive support planning can be improved by using the admittance and estimated time to recovery in conjunction with the voltage for FIDVR quantification.

## VII. REFERENCES

[1] DOE-NERC FIDVR Conf., Sep. 29, 2009. [Online]. Available: http://www.nerc.com/files/FIDVR-Conference-Presentations-9-29-09.pdf.
[2] B. R. Williams, W. R. Schmus, and D. C. Dawson, "Transmission voltage recovery delayed by stalled air conditioner compressors," IEEE Trans. Power Syst., vol. 7, no. 3, pp. 1173–1181, Aug. 1992.
[3] North American Transmission Forum, Transient voltage criteria reference document, September 2016. [online]
[4] V. Krishnan, H. Liu, and J. D. McCalley, "Coordinated reactive power planning against power system voltage instability," in Proc. IEEE/PES Power Systems Conf. Expo., 2009 (PSCE '09), Mar. 2009, pp. 1–8
[5] B. Sapkota and V. Vittal, "Dynamic var planning in a large power system using trajectory sensitivities," IEEE Trans. Power Syst., vol. 25, no. 1, pp. 461–469, Feb. 2010.
[6] M. Paramasivam, S. Dasgupta, V. Ajjarapu and U. Vaidya, "Contingency Analysis and Identification of Dynamic Voltage Control Areas," in IEEE Transactions on Power Systems, vol. 30, no. 6, pp. 2974-2983, Nov. 2015.
[7] S. M. Halpin, et. al, "Slope permissive under-voltage load shed relay for delayed voltage recovery mitigation," IEEE Trans. Power Syst., vol. 23, no. 3, pp. 1211–1216, Aug. 2008.
[8] S. M. Halpin, et. al, "The MVA-volt index: A screening tool for predicting fault-induced low voltage problems on bulk transmission systems," IEEE Trans. Power Syst., vol. 23, no. 3, pp. 1205–1210, Aug. 2008.
[9] S. V. Kolluri, et. al, "Relay-based undervoltage load shedding scheme for Entergy's Western Region," 2015 IEEE Power & Energy Society General Meeting, Denver, CO, 2015, pp. 1-5.
[10] H. Bai, and V. Ajjarapu "A novel online load shedding strategy for mitigating fault-induced delayed voltage recovery" IEEE Trans. Power Syst. vol. 26 no. 1 pp. 294-304 Feb. 2011.
[11] A. Mahari and H. Seyedi, "A fast online load shedding method for mitigating FIDVR based on novel stability index," ICEE-2013, pp. 1-6.
[12] WECC MVWG Dynamic Composite Load Model Specifications (Jan 2015) [online].
[13] Qiuhua Huang, et. al. "A Reference Implementation of WECC Composite Load Model in Matlab and GridPACK", [online] arXiv:1708.00939.
[14] H. Zheng and C. L. DeMarco, "A New Dynamic Performance Model of Motor Stalling and FIDVR for Smart Grid Monitoring/Planning," in IEEE Transactions on Smart Grid, vol. 7, no. 4, pp. 1989-1996, July 2016.
[15] Chassin D.P., et al, "Load Modeling Activities", Report-24425 by PNNL. http://www.pnnl.gov/main/publications/external/technical_reports/PNNL-24425.pdf
[16] Load Model Data Tool by Pacific Northwest National Laboratory, [online] https://svn.pnl.gov/LoadTool.
[17] M.W. Tenza, et. al., "An Analysis of the Sensitivity of WECC Grid Planning Models to Assumptions Regarding the Composition of Loads", Mitsubishi Electric Power Products, Nov 2016.
[18] M. Paramasivam, " Dynamic optimization based reactive power planning for improving short-term voltage performance", PhD Thesis, [online] https://lib.dr.iastate.edu/etd/14945/
[19] S. Robles, "2014 FIDVR Events Analysis on Valley Distribution Circuits". Prepared for LBNL by Southern California Edison, 2015
[20] R. J. Bravo, "DER volt-VAr and voltage ride-through needs to contain the spread of FIDVR events," 2015 IEEE PESGM, Denver, 2015, pp. 1-3.

## VIII. BIOGRAPHIES

**Amarsagar Reddy Ramapuram Matavalam,** (S'13) received the B.Tech. degree in Electrical Engineering and the M.Tech. degree in Power Electronics and Power Systems, both from IIT-M, Chennai, India. He is currently pursuing Ph.D. in the Department of Electrical and Computer Engineering at Iowa State University, Ames, IA, USA. His research is in voltage stability analysis, power systems data analytics, and numerical techniques for power systems.

**Venkataramana Ajjarapu** (S'86, M'86, SM'91, F'07) received the Ph.D. degree in electrical engineering from the University of Waterloo, Waterloo, ON, Canada, in 1986. Currently, he is a Professor in the Department of Electrical and Computer Engineering at Iowa State University, Ames, IA, USA. His research is in voltage stability analysis and nonlinear voltage phenomena.